\documentclass[%
 reprint,
 amsmath,amssymb,
 aps,
 prl,
]{revtex4-2}

\usepackage{pdfpages} 
\usepackage{pgffor} 

\makeatletter
\AtBeginDocument{\let\LS@rot\@undefined}
\makeatother

\def\supplementfilename{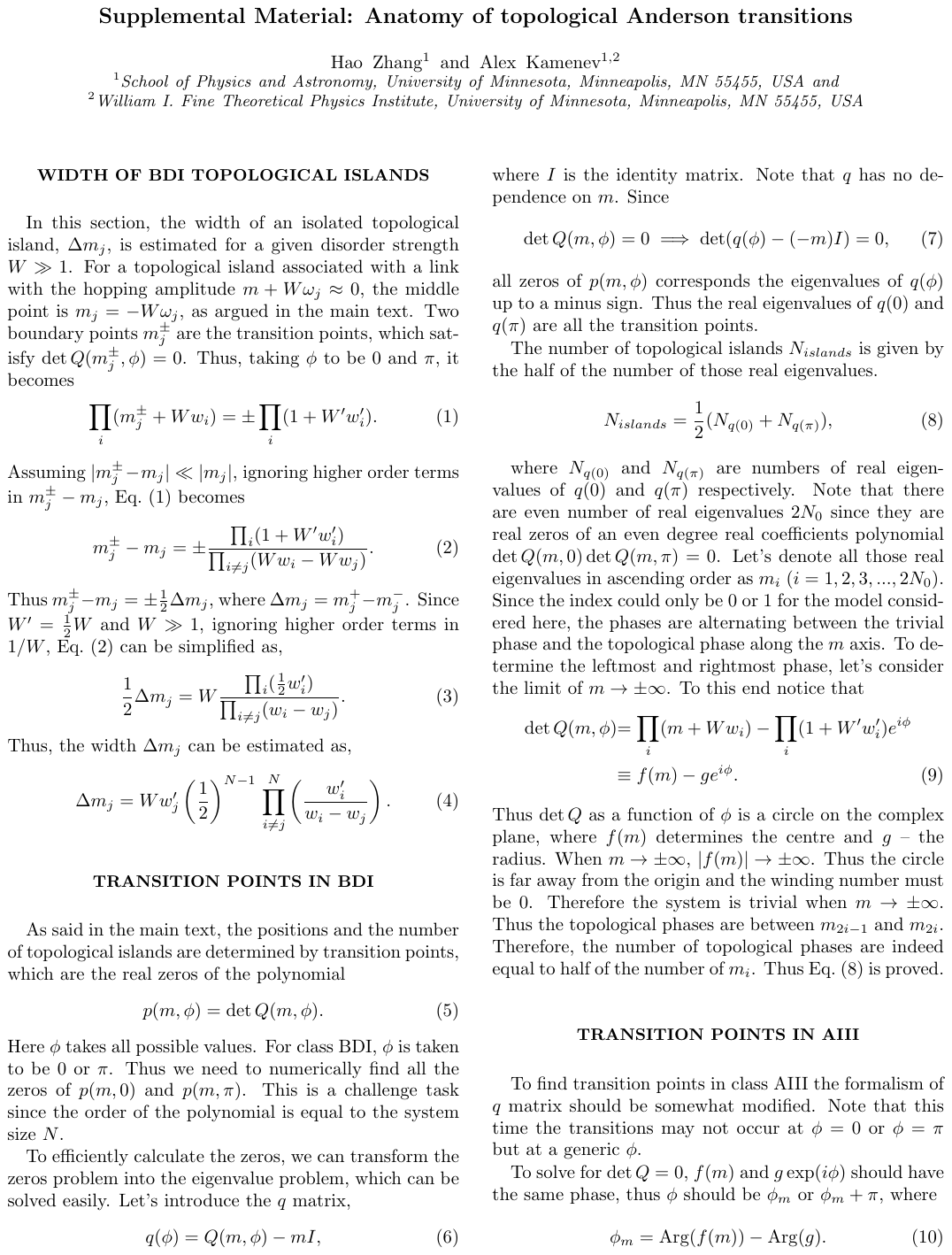}

\pdfximage{\supplementfilename}
\def\numbersupplementpages{\the\pdflastximagepages}

\newif\ifarXiv
\arXivtrue 

\usepackage{graphicx}
\usepackage{dcolumn}
\usepackage{bm}

\usepackage{subcaption}
\usepackage{xcolor}

\begin{document}

\preprint{APS/123-QED}

\title{Anatomy of topological Anderson transitions}

\author{Hao Zhang$^{1}$ and Alex Kamenev$^{1,2}$}

\affiliation{$^1$School of Physics and Astronomy, University of Minnesota, Minneapolis, MN 55455, USA}
\affiliation{$^2$William I. Fine Theoretical Physics Institute, University of Minnesota, Minneapolis, MN 55455, USA}

\date{\today}
\vspace{0.1cm}

\begin{abstract}
We study mesoscopic signatures of the topological Anderson transitions in topological disordered chains.
To this end we introduce an integer-valued sample-specific definition of the topological index in finite size systems.  Its phase diagram exhibits a fascinating structure of intermittent topological phases, dubbed topological islands. Their existence is rooted in the real zeros of the underlying random polynomial.  Their statistics exhibits finite-size scaling, pointing to the location of the bulk topological Anderson transition. While the average theories in AIII and BDI symmetry classes are rather similar, the corresponding patterns of topological islands and their statistics are qualitatively different.  We also discuss observable signatures of sharp topological transitions in mesoscopic systems, such as persistent currents and entanglement spectra. 
\end{abstract}

\maketitle


An interplay between topology and disorder plays an essential role in our understanding of topological materials \cite{bernevig_topological_2013,asboth_short_2016,shen_topological_2017,qi_topological_2011,hasan_colloquium:_2010,chiu_classification_2016}. Most notably it leads to a concept of topological Anderson insulators \cite{li_topological_2009,altland_quantum_2014,altland_topology_2015,prodan_entanglement_2010,mondragon-shem_topological_2014,meier_observation_2018,stutzer_photonic_2018,groth_theory_2009,vu_weak_2022,loring_disordered_2010,jiang_numerical_2009,prodan_disordered_2011,xing_topological_2011,song_aiii_2014,fulga_scattering_2011,haim_benefits_2019,hua_disorder-induced_2019,akhmerov_quantized_2011,bardarson_aharonov-bohm_2010,li_topological_2020,claes_disorder_2020,wauters_localization_2019,hsu_disorder-induced_2021,velury_topological_2021,liu_real-space_2021,antonenko_mesoscopic_2020,titum_anomalous_2016,titum_disorder-induced_2015,guo_topological_2010,kobayashi_disordered_2013,jiang_stabilizing_2012,garcia_real-space_2015,yamakage_disorder-induced_2011,song_dependence_2012,ryu_disorder-induced_2012,zhang_non-hermitian_2020,zhang_localization_2012,yang_higher-order_2021,shi_disorder-induced_2021}, where the robustness of the topological index is protected by the localized nature of wave-functions, rather than a band gap in the energy spectrum. Phase diagrams of topological Anderson insulators generically exhibit transitions between localized phases with distinct topological indexes. Remarkably, localization length diverges at such transitions \cite{song_aiii_2014,mondragon-shem_topological_2014}, indicating the presence of extended states.  This happens even in 1d, where the {\em ensemble averaged}  theory in the thermodynamic limit is understood  in terms of a two parameters scaling \cite{altland_quantum_2014,altland_topology_2015}, similar to  2d integer quantum Hall effect \cite{PRUISKEN1984}.

Although conceptually appealing, the average theory misses a trove of interesting information regarding sample-specific properties of disordered systems. The main goal of this paper is to uncover  a fascinating hidden structure of intermittent topological phases, dubbed {\em topological islands}. Both the number of such  islands and their locations on the phase diagram are determined by the real zeros of the underlying random polynomial and are completely lost in any ensemble averaged treatment. A key step in this direction is a physically meaningful definition of an integer-valued  topological index, which does not rely  
on the ensemble averaging, nor on the thermodynamic limit, cf. \cite{prodan_entanglement_2010,prodan_disordered_2011,song_aiii_2014,mondragon-shem_topological_2014}. 

We show that the finite-size scaling of the number of topological islands is a sensitive tool to reveal locations of the bulk topological Anderson transitions. 
Moreover, such finite-size scaling looks qualitatively distinct in different symmetry classes, eg., AIII vs. BDI, while their ensemble-average descriptions \cite{altland_topology_2015} are very much alike.  

Our results are important for ongoing experimental efforts to detect 1D topological phase transitions and localization-delocalization phenomena. In many real experiments, such as recent topological Anderson insulator observations\cite{meier_observation_2018}, the system size is typically not large. As a result, self-averaged theories are less suitable for analysis. However, our results demonstrates what features one may expect in a real experiment setup, even when dealing with smaller systems. In the end, we discuss observable manifestations of topological transitions in  mesoscopic size disordered samples\cite{meier_observation_2018, barkhofen_experimental_2023} on persistent currents as well as entanglement spectra.

\begin{figure}[h!]
\centering

\begin{subfigure}{.475\textwidth}
	\centering
	\includegraphics[width=\linewidth]{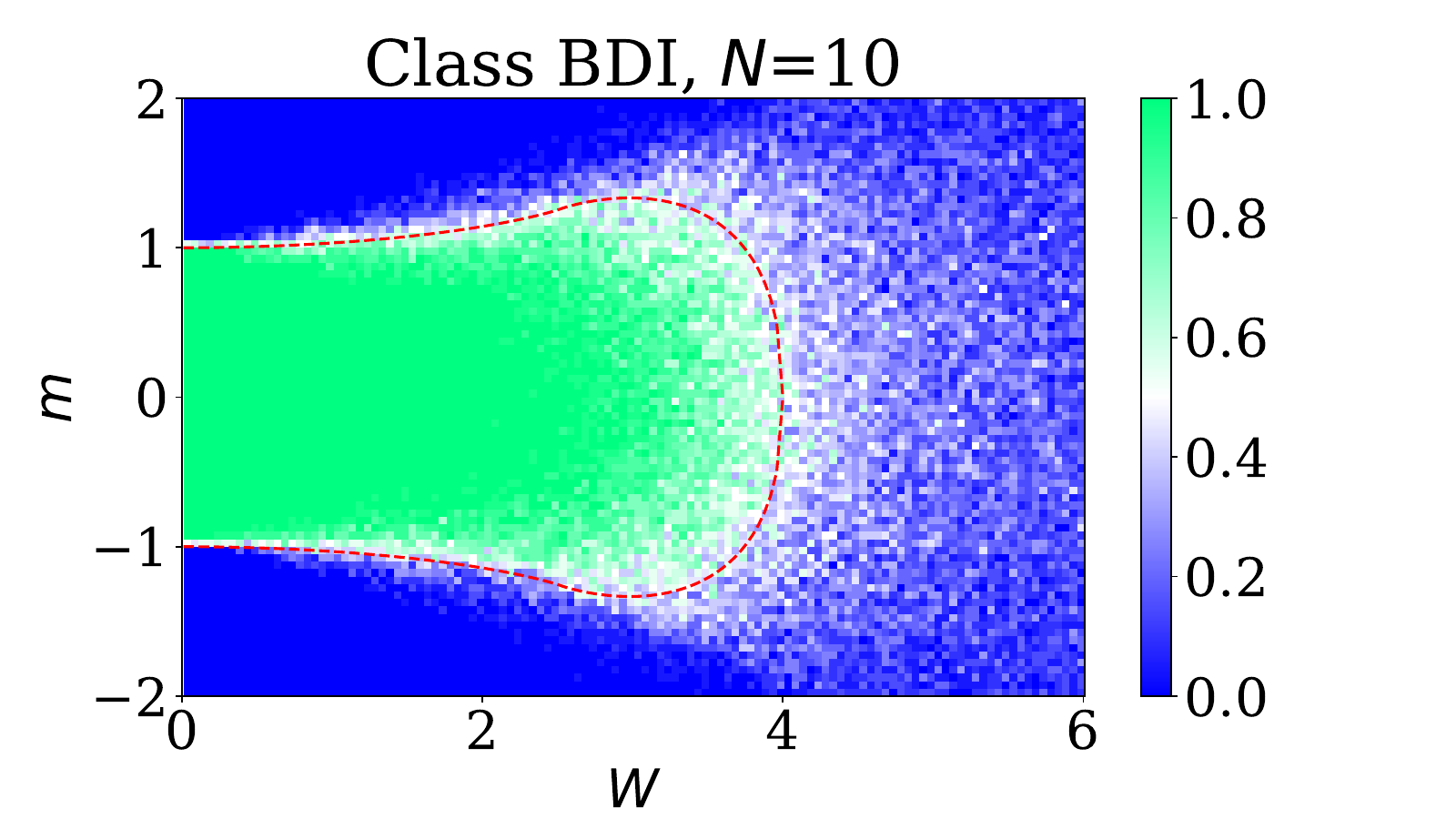}
	\caption{}
	\label{fig:phase_diagram_average_index_BDI}
\end{subfigure}

\begin{subfigure}{.475\textwidth}
	\centering
	\includegraphics[width=\linewidth]{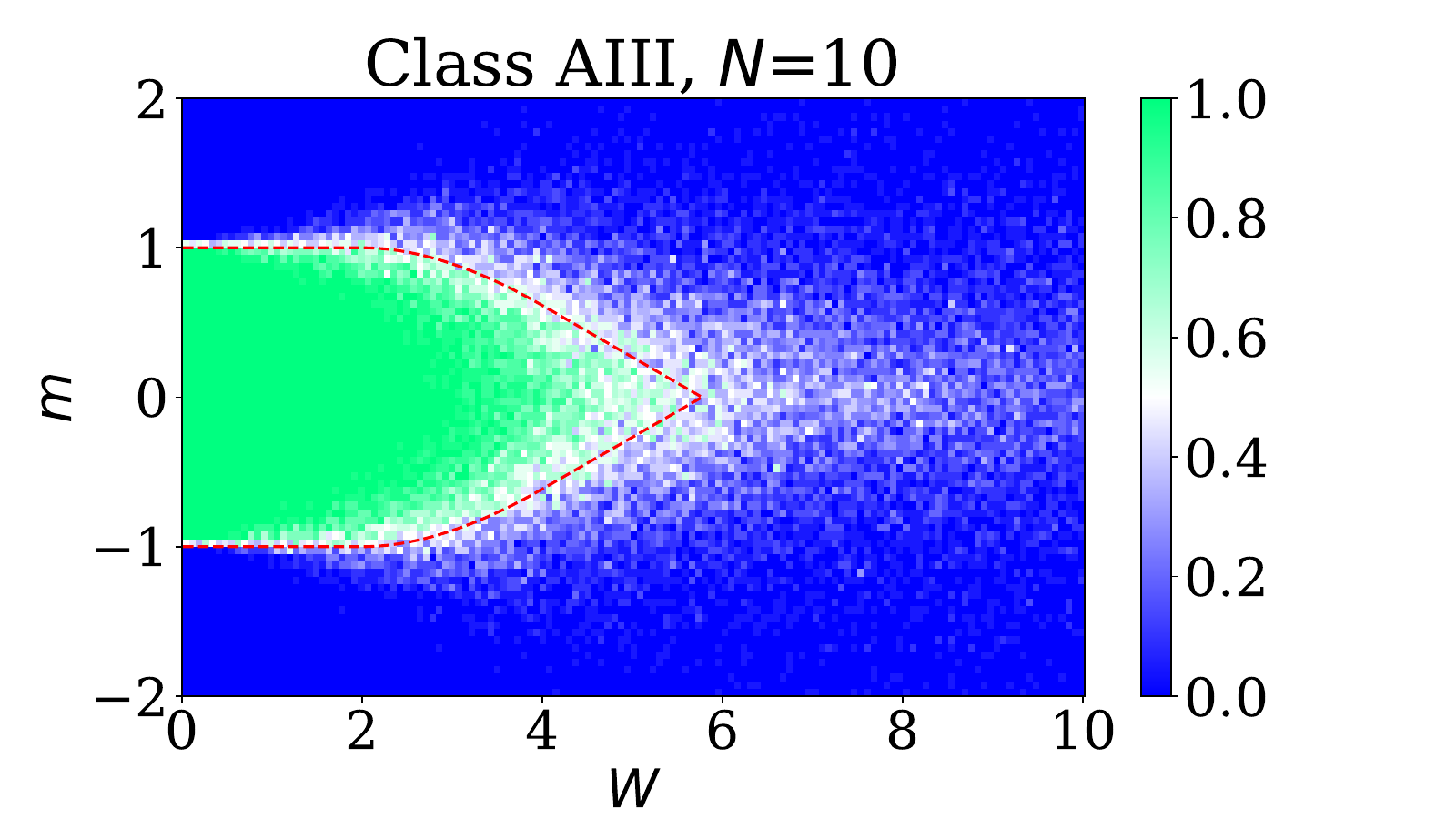}
	\caption{}
	\label{fig:phase_diagram_average_index_AIII}
\end{subfigure}

\caption{Average topological index on the $(W,m)$ phase diagram for $N=10$. To emphasize fluctuations, each point is averaged using a small sample size of $20$. The red lines are the phase boundaries of the corresponding infinite systems. (a) Class BDI. (b) Class AIII.}
\label{fig:phase_diagram}
\end{figure}

To illustrate the idea, let's consider a finite size disordered system defined on a ring, 
\begin{equation}\label{eq:Hamiltonian}
    H=\sum_{j=1}^N t_{j}c_{j,A}^\dagger c_{j,B}+t_{j}'c_{j,B}^\dagger c_{j+1,A}+h.c.,
\end{equation}
where $c_{N+1,A}=c_{1,A}$. Here A,B label two sub-lattices and $t_{j}$, $t_{j}'$ are hopping strengths within the unit cell and between nearest unit cells, respectively. The chiral symmetry preserving disorder is introduced as,
\begin{eqnarray}
t_{j} = m +Ww_{j}; \qquad \qquad
t_{j}'=1+W'w_{j}',
\end{eqnarray}
where $m$ is a uniform staggering, $W,W'$ are the disorder strengths and $w_{j},w_{j}'$ are independent random variables uniformly drawn from the box $[-1/2,1/2]$. In all subsequent examples $W'=\frac{1}{2}W$. The system possesses the chiral symmetry,
\begin{equation}
    \tau_z H \tau_z=-H,
\end{equation}
where $\tau_z$ is the Pauli matrix acting in the sub-lattices space. In the time reversal symmetric BDI class \cite{altland_nonstandard_1997,chiu_classification_2016} all the parameters are real, while generalization to the AIII class is discussed below.

The usual topological index in the $k$ space cannot be defined in a finite-size and translationally non-invariant system. To overcome it we introduce  a  flux, threading the ring \cite{altland_quantum_2014, fulga_topological_2011},  through the substitution  
\begin{equation}
    t_Nc^\dagger_{N,B}c_{1,A} \to t_Ne^{i\phi}c^\dagger_{N,B}c_{1,A}.
\end{equation}
The energy spectrum consists of $2N$ particle-hole 
symmetric ``bands'' as functions of flux $\phi\in[0,2\pi]$. One can now introduce the index by using the Hamiltonian in a chiral basis
\begin{equation}
  H(m,\phi)=\begin{pmatrix}0&Q(m,\phi)\\Q^\dagger(m,\phi) &0\end{pmatrix},
\end{equation}
and defining 
\begin{equation}\label{eq: theIndex}
    \nu=\frac{1}{2\pi i}\int_0^{2\pi}\!\!\! d\phi\,\,   \partial_\phi\log\det Q(m,\phi)
\end{equation}
Since this is a winding number of the $\phi\mapsto\det Q(m,\phi)$ map, the index is integer-valued.

A transition point, $m=m_c$, is marked by the gap at zero energy closing for some $\phi=\phi_{m_c}$. At this instance there are two zero energy eigenvalues which implies $\det H=-|\det Q|^2=0$. This means that, as $\phi$ goes from $0$ to $2\pi$,  $\det Q(m_c,\phi)$ draws a closed loop in the complex plane which passes through the origin. Its winding number is thus undefined, while for $m\neq m_c$ it is an integer, which jumps by one over the transition.  For the BDI class,  $\phi_{m_c}$ is either $0$ or $\pi$ due to the time reversal symmetry. 

\begin{figure}[htbp]
	\centering
	\includegraphics[width=\linewidth]{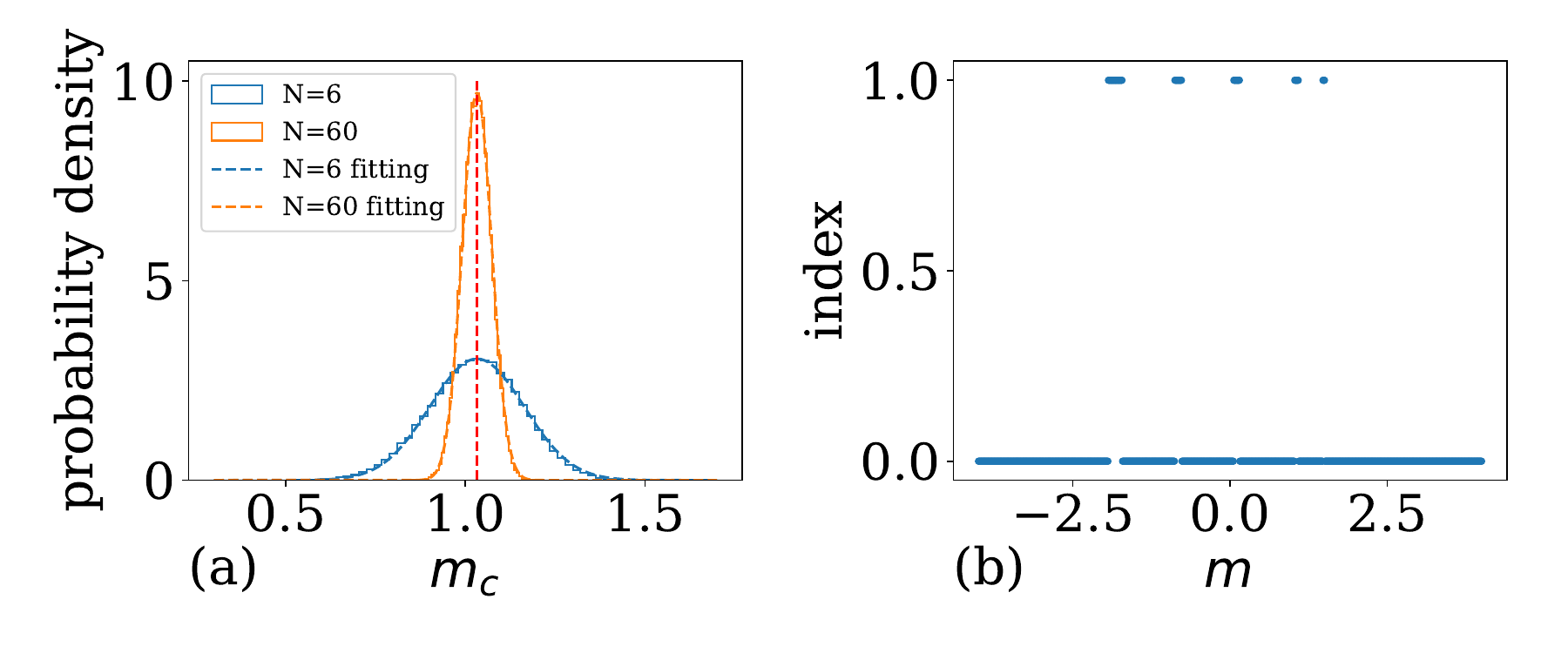}
	\caption{(a) The distribution over $10^5$ realisations of the transition points near $m=1$ for $W=1$; $N=6$ and $N=60$ are shown in blue and orange along with the Gaussian fits.  (b) Topological  index for a system with $N=6, W=4$ and a fixed disorder realization. Notice five intermittent topological phases, dubbed topological islands.}
	\label{fig:locandislands}
\end{figure}
 
 Figure~\ref{fig:phase_diagram} shows the topological index for N=10, averaged over 20 disorder realizations, on the phase plane of $(W,m)$.  The red line is the topological phase boundary of the corresponding infinite system, which is calculated using the method of Ref.~[\onlinecite{mondragon-shem_topological_2014}]. The enhanced fluctuations can be seen around the boundary. Notice that, in a certain region, they extend far beyond the bulk boundary to a very strong  disorder. 

In the weak disorder region, each realization exhibits  two transition points located near $m\approx \pm 1$. 
The sample-to-sample fluctuations of each of these transition points,  Fig.~\ref{fig:locandislands}(a), are well fitted by the Gaussian distribution. Its center follows the red boundary, while the width shrinks as $N\to\infty$. This marks a self-averaging transition.  

However, when the disorder is strong, the fluctuations persist even in the $N\to\infty$ limit. This is attributed to the hidden structure of topological islands -- the intermittent topological phases. Figure.~\ref{fig:locandislands}(b) shows a sample-specific topological index as a function of $m$ for $W=4$.  The number of topological islands can be as large as $N$. They do not disappear in the $N\to\infty$ limit but their width and relative area scales to zero.  

\begin{figure*}[htbp]
	\centering
	\includegraphics[width=\linewidth]{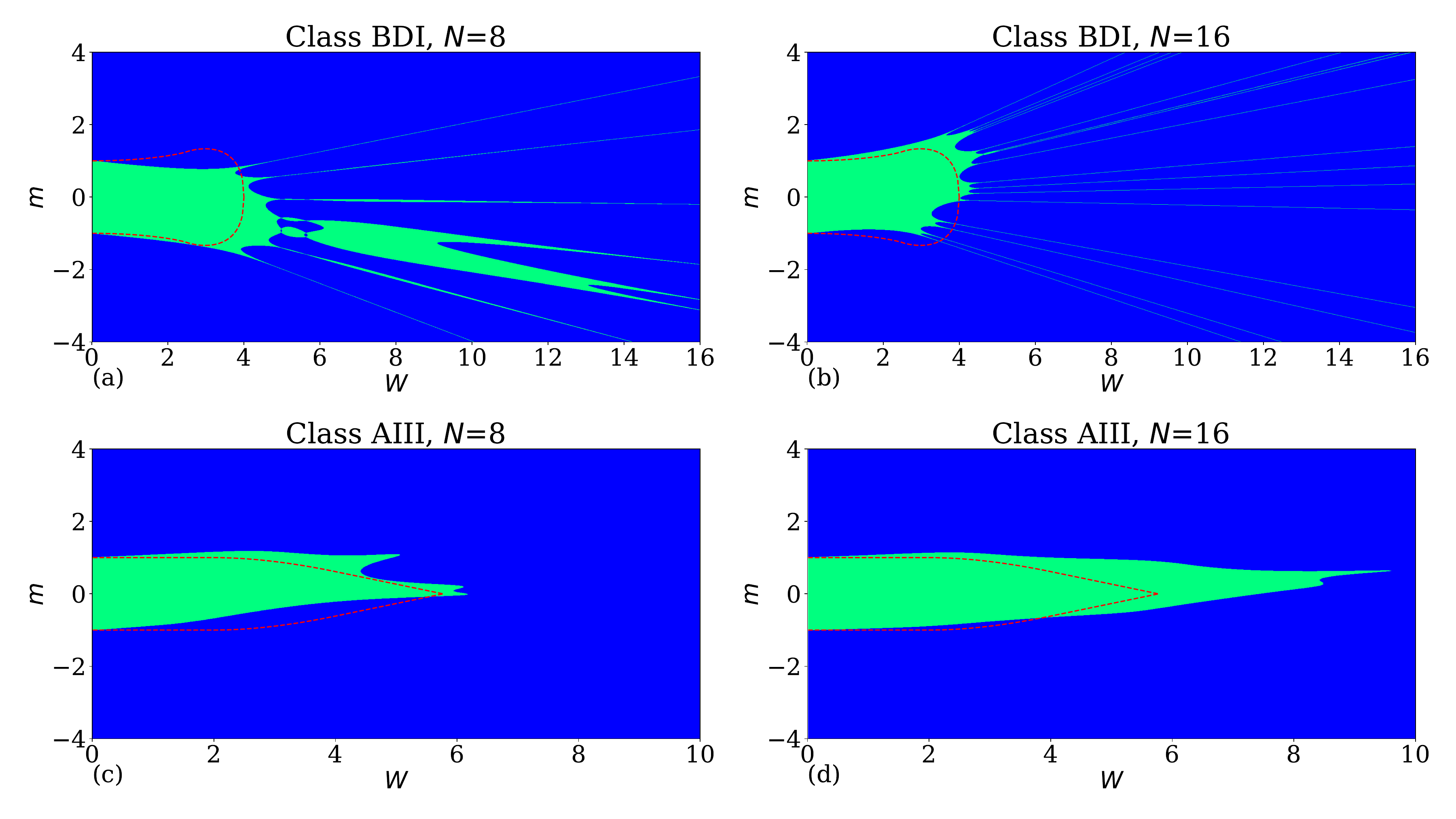}
	\caption{Phase diagrams of  fixed disorder realizations. The green region is $\nu=1$, while the blue region is $\nu=0$. The red lines are the phase boundaries of the corresponding infinite systems \cite{mondragon-shem_topological_2014}. (a), (b)  Class BDI with $N=8$, $N=16$.  (c), (d) Class AIII with $N=8$, $N=16$. }
	\label{fig:stripes}
\end{figure*}

For a better view of topological islands, the phase diagrams for fixed disorder realizations are shown in Fig.~\ref{fig:stripes}. The disorder realizations are fixed except the overall amplitude, $W$. The green region is topological, while the blue one is trivial. The red lines are the phase boundary of the corresponding infinite systems. The bulk topological region of the BDI model grows into $N$ topological islands.  These islands extend to an arbitrarily strong disorder. Towards a weaker disorder they thicken and coalesce, ultimately  forming the bulk topological region. 

The presence of topological islands is a general feature, not restricted to the particular model we examined. They are determined by the real zeros of the random polynomial

\begin{equation}
    p(m,\phi)=\det Q(m,\phi),
\end{equation}
Here, $\phi$ takes all possible values. Those zeros are the boundaries of the islands and can be used to determine the number and locations of them. Topological islands exists as long as there are many real zeros of $p(m,\phi)$. Let's consider a general random BDI model with all ranges of longer hopping. We first introduce the control parameter $m$ as the hopping within the unit cell and thus m is the diagonal element of $Q$. Then we introduce random hopping between different sites while respecting the chiral symmetry. $Q$ may be modeled as $mI + R$, where $I$ is the identity matrix and $R$ is a random matrix. Real zeros of $p(m,\phi)$ are thus the same as real eigenvalues of $R$ (up to a minus sign). There are around $\sqrt{N}$ real eigenvalues of $R$, where $N$ is the number of unit cells.\cite{edelman_how_1994} For class BDI, $\phi$ takes $0$ and $\pi$, which leads to $2\sqrt{N}$ transitions. Hence, we expect there are roughly $\sqrt{N}$ topological islands for a generic BDI model.

For the specific model we consider, real zeros can be found by requiring  the product of the absolute value of all intra unit cell hopping strength equals that of inter unit cell hopping strength
 $\prod\limits_{j=1}^N \left|t_j\right| = \prod\limits_{j=1}^N \left|t_j'\right|.$ 
 In the bulk of $\nu=0$ (blue) region, $\prod_j \left|t_j\right|>\prod_j \left|t_j'\right|$. 
The radiating shape of topological islands can be understood by looking at the special points where the chain is accidentally cut in two, given that one of $t_j=0,$ 
\begin{equation} \label{eq:topological_line}
    m+W w_j=0.
\end{equation}
Real zeros of the random polynomial are usually near them for large $W$. These straight lines on the $(W,m)$ plane, with a set of random slopes $-w_j$,   mark the centers of the islands.  The width of $\nu=1$ (green) region around each
such a line  may be estimated for $W\gg1$ as \footnote{See Supplemental Material at [URL will be inserted by publisher] for the derivation.},
\begin{equation}
    \Delta m_j = W w_j' \left(\frac{1}{2}\right)^{N-1}\, 
    \prod_{i\neq j}^N \left(\frac{w_i'}{w_i-w_j}\right).
\end{equation}
Thus the angular width of the island, $\Delta m_j/W$, is a fixed number for a given realization. It decreases exponentially with $N\to\infty$.

\begin{figure*}[htbp]
	\includegraphics[width=\linewidth]{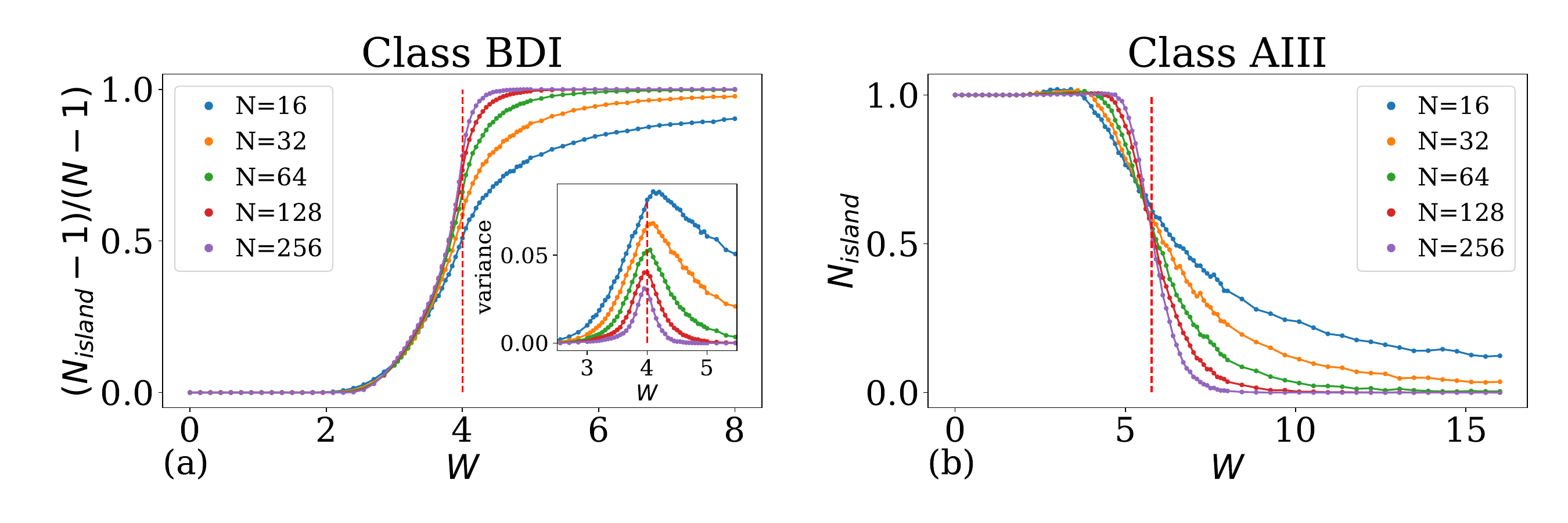}
	\caption{The average number of islands versus the disorder strength $W$ for systems with different system sizes $N$. (a) Class BDI, the convenient quantity is  $\frac{N_{islands}-1}{N-1}.$  Inset shows its variance, which peaks at $W_c=4$ as $N\to\infty$. (b) Class AIII,  the graphs shows a crossing point at $W_c=4/\log(2)$.}
	\label{fig:num_islands}
\end{figure*}

To further understand the anatomy of the topological islands, we look at the number of islands \footnote{See Supplemental Material at [URL will be inserted by publisher] for the details of numerical calculations.} across all $m$'s for a fixed $W$, Fig.~\ref{fig:num_islands}(a).  A convenient way to represent data is to plot $\frac{N_{islands}-1}{N-1}$, vs. $W$.  As $N$ increases it approaches a limiting function, smoothly interpolating between zero and one. To detect the exact location of the transition we look for  the variance of the number of islands (divided by ${N-1}$), see the inset. It shows that as $N$ increases, the variance peak become more narrow and centered  at $W_c=4$. This illustrates that the bulk transition point for $m=0$ may be identified as the maximum of the variance of the number of islands in a finite size simulation.

We turn now to the AIII symmetry class with the broken time-reversal symmetry. To this end we use Hamiltonian 
 Eq.~\ref{eq:Hamiltonian} with  the complex hopping amplitudes. Namely, the absolute values and phases of $w_j$ and $w_j'$ are now uniformly drawn from $(0,1/2$) and $(-\pi,\pi)$ respectively. Figure~\ref{fig:phase_diagram}(b) shows the average phase diagram of class AIII. It is similar to that of BDI class and so is the average theory of the corresponding bulk topological Anderson transition \cite{altland_quantum_2014,altland_topology_2015,song_aiii_2014,mondragon-shem_topological_2014}. The latter takes place along the red line, where the localization length diverges.

However, the sample-specific fluctuations of the topological index behaves in a way, which is very different from the BDI class. Indeed,  the topological islands now have a finite extent 
and are limited to a relatively weak disorder part of the phase diagram,  Fig.\ref{fig:stripes}(c)(d).  Moreover, there are  only few  islands and their number does not increase with $N$. This is due to complex random hopping amplitudes which statistically exclude instances of a cut chain (real and imaginary parts do not vanish at the same $W$).   The average number of islands across all $m$'s is plotted in Fig.~\ref{fig:num_islands}(b) for different system sizes $N$. It shows a well-defined crossing point at $W_c=4/\log(2)$, 
in agreement with the bulk transition point at $m=0$ \cite{mondragon-shem_topological_2014}.
As $N$ grows the number of islands approaches 
one and zero for $W<W_c$ and $W>W_c$, correspondingly.

We have shown that mesoscopic sample-to-sample fluctuations manifest themselves in the formation of topological islands. Their number, shape and statistics are qualitatively different between BDI and AIII classes. However, in both cases their finite size scaling provides the exact location of the corresponding bulk topological Anderson transition.

Before concluding we discuss observable signatures of the {\em sharp} topological transitions in finite size systems.

\begin{figure}[htbp]
	\centering
	\includegraphics[width=\linewidth]{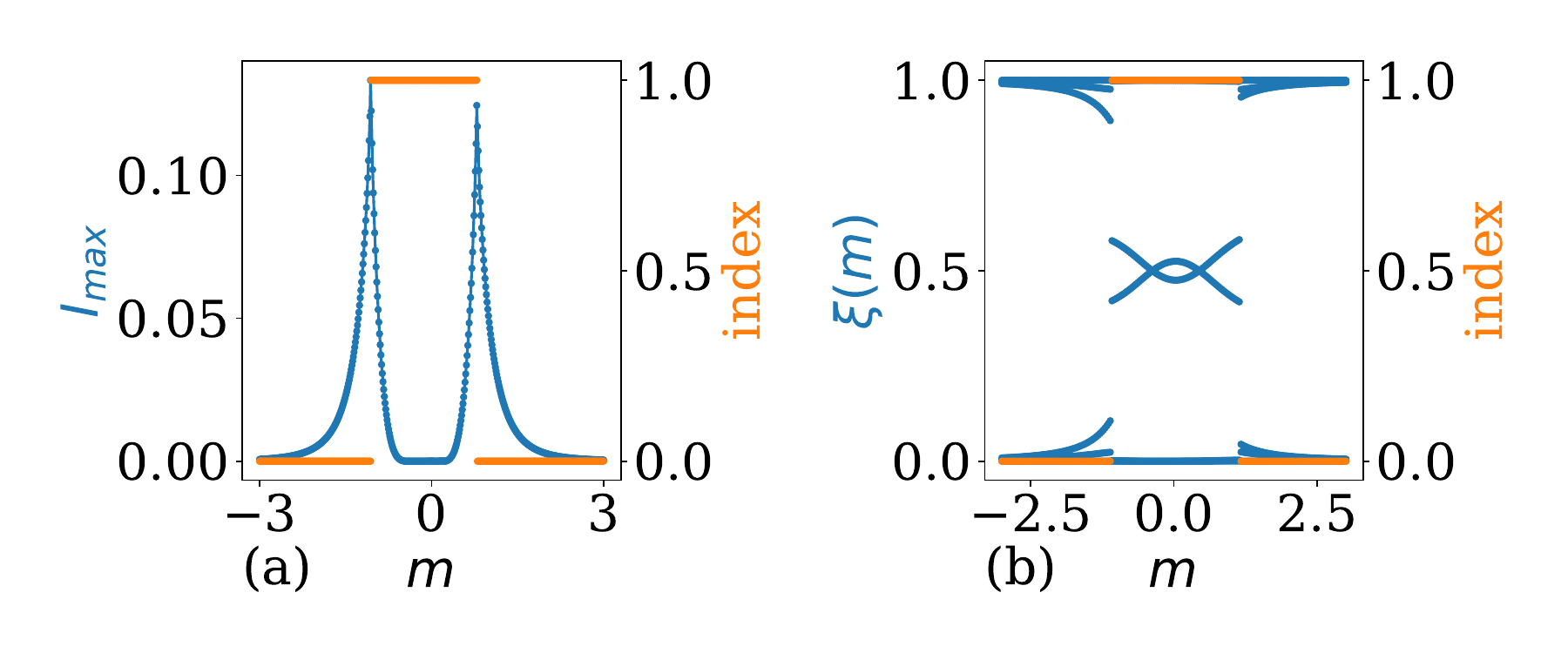}
	\caption{(a) Sample-specific maximum value of the persistent current, $I_\text{max}$, (blue dots) and the topological index, $\nu$, (orange dots) as  functions of $m$.   $I_\text{max}$ peaks at the two transition points with a power law singularities. (b) The entanglement spectrum of a BDI realization with $N=6$ and $W=1$. The blue dots are $\xi_i$, which are related to the entanglement energies through Eq.(\ref{eq:xi}). The orange dots are the topological index.  The entanglement spectrum has discontinuities at transitions. In the topological phase, there are two $\xi_i$ around $0.5$ (and thus $\epsilon_i$ around 0). }
	\label{fig:singatures}
\end{figure}

First, consider a mesoscopic persistent current, given by \cite{bleszynski-jayich_persistent_2009,ghosh_persistent_2014,nava_persistent_2017,sticlet_persistent_2013}
\begin{equation}
    I(m,\phi) = -\partial_\phi E(m,\phi),
\end{equation}
where $E(m,\phi)$ is the ground state energy of the half-filled system. It is a periodic function $\phi$, which exhibits a maximum at a certain $\phi$,  
$I_\text{max}(m)=\max_{\phi}I(m,\phi)$. 
It appears that this maximal value exhibits a non-analytic maximum at the topological transition critical $m_c$,  Fig.~\ref{fig:singatures}(a), 
\begin{equation}
    I_\text{max} \propto -|m-m_c|^\alpha,
\end{equation}
where  $\alpha$ is the critical exponent that may have dependence on system parameters.

In simple models with nearest neighbor hopping, there are zero energy states located close to weak links in topological phases and they resemble the edge states while in more general models, their localization can’t be tight to any visible edge.

The transitions also have implications in entanglement spectra \cite{li_entanglement_2008,prodan_entanglement_2010}. We divide the chain into two equal parts A and B (do not confuse with the sub-lattices) and calculate the trace over the B part.  The corresponding reduced density matrix of A takes the form  
\begin{equation}
    \rho_A=\sum_{n_B}\langle n_B |\rho|n_B\rangle \propto e^{-\sum_i \epsilon_i a_i^\dagger a_i},
\end{equation}
where $|n_B\rangle$ are basis vectors of the B part and $a_i$ are normal modes.
The second equality is a property of  Gaussian (i.e. non-interacting) models \cite{peschel_calculation_2003}.  Here $\epsilon_i$ is the entanglement spectrum, which may be expressed through  the eigenvalues $\xi_i$ of the one-particle covariance matrix $C_{jj'}=\langle c_j^\dagger c_{j'}\rangle$ \cite{peschel_calculation_2003} as
\begin{equation} \label{eq:xi}
    \epsilon_i = \frac{1}{2}\log\left(\frac{1-\xi_i}{\xi_i}\right),
\end{equation}
where $j$ and $j'$ label the lattice sites of the A part and $\langle\ldots\rangle$ denote ground state averaging.
The entanglement spectrum exhibits  discontinuities at topological transitions, see Fig.~\ref{fig:singatures}(b). For a weak disorder, the spectrum also exhibits nearly zero entanglement energies within the topological phase. It can thus be considered as a means to identify the sharp topological transitions in mesoscopic systems. 




We are grateful to Xuzhe Ying for useful discussions. This work was supported  by the NSF under Grant No. DMR-2037654. 

\begin{acknowledgments}
\end{acknowledgments}


\bibliography{main}

\ifarXiv
    

\section*{Supplementary material}

# Supplemental Material: Anatomy of topological Anderson transitions

Hao Zhang[1] and Alex Kamenev[1,2]
[1]*School of Physics and Astronomy, University of Minnesota, Minneapolis, MN 55455, USA* and
[2]*William I. Fine Theoretical Physics Institute, University of Minnesota, Minneapolis, MN 55455, USA*

## WIDTH OF BDI TOPOLOGICAL ISLANDS

In this section, the width of an isolated topological island, $\Delta m_j$, is estimated for a given disorder strength $W \gg 1$. For a topological island associated with a link with the hopping amplitude $m + W\omega_j \approx 0$, the middle point is $m_j = -W\omega_j$, as argued in the main text. Two boundary points $m_j^\pm$ are the transition points, which satisfy $\det Q(m_j^\pm, \phi) = 0$. Thus, taking $\phi$ to be 0 and $\pi$, it becomes

$$\prod_i (m_j^\pm + W w_i) = \pm \prod_i (1 + W' w_i'). \tag{1}$$

Assuming $|m_j^\pm - m_j| \ll |m_j|$, ignoring higher order terms in $m_j^\pm - m_j$, Eq. (1) becomes

$$m_j^\pm - m_j = \pm \frac{\prod_i (1 + W' w_i')}{\prod_{i \neq j} (W w_i - W w_j)}. \tag{2}$$

Thus $m_j^\pm - m_j = \pm \frac{1}{2}\Delta m_j$, where $\Delta m_j = m_j^+ - m_j^-$. Since $W' = \frac{1}{2}W$ and $W \gg 1$, ignoring higher order terms in $1/W$, Eq. (2) can be simplified as,

$$\frac{1}{2}\Delta m_j = W \frac{\prod_i (\frac{1}{2} w_i')}{\prod_{i \neq j}(w_i - w_j)}. \tag{3}$$

Thus, the width $\Delta m_j$ can be estimated as,

$$\Delta m_j = W w_j' \left(\frac{1}{2}\right)^{N-1} \prod_{i \neq j}^N \left(\frac{w_i'}{w_i - w_j}\right). \tag{4}$$

## TRANSITION POINTS IN BDI

As said in the main text, the positions and the number of topological islands are determined by transition points, which are the real zeros of the polynomial

$$p(m, \phi) = \det Q(m, \phi). \tag{5}$$

Here $\phi$ takes all possible values. For class BDI, $\phi$ is taken to be 0 or $\pi$. Thus we need to numerically find all the zeros of $p(m, 0)$ and $p(m, \pi)$. This is a challenge task since the order of the polynomial is equal to the system size $N$.

To efficiently calculate the zeros, we can transform the zeros problem into the eigenvalue problem, which can be solved easily. Let's introduce the $q$ matrix,

$$q(\phi) = Q(m, \phi) - mI, \tag{6}$$

where $I$ is the identity matrix. Note that $q$ has no dependence on $m$. Since

$$\det Q(m, \phi) = 0 \implies \det(q(\phi) - (-m)I) = 0, \tag{7}$$

all zeros of $p(m, \phi)$ corresponds the eigenvalues of $q(\phi)$ up to a minus sign. Thus the real eigenvalues of $q(0)$ and $q(\pi)$ are all the transition points.

The number of topological islands $N_{islands}$ is given by the half of the number of those real eigenvalues.

$$N_{islands} = \frac{1}{2}(N_{q(0)} + N_{q(\pi)}), \tag{8}$$

where $N_{q(0)}$ and $N_{q(\pi)}$ are numbers of real eigenvalues of $q(0)$ and $q(\pi)$ respectively. Note that there are even number of real eigenvalues $2N_0$ since they are real zeros of an even degree real coefficients polynomial $\det Q(m,0) \det Q(m,\pi) = 0$. Let's denote all those real eigenvalues in ascending order as $m_i$ ($i = 1, 2, 3, ..., 2N_0$). Since the index could only be 0 or 1 for the model considered here, the phases are alternating between the trivial phase and the topological phase along the $m$ axis. To determine the leftmost and rightmost phase, let's consider the limit of $m \to \pm\infty$. To this end notice that

$$\det Q(m, \phi) = \prod_i (m + W w_i) - \prod_i (1 + W' w_i') e^{i\phi}$$
$$\equiv f(m) - g e^{i\phi}. \tag{9}$$

Thus $\det Q$ as a function of $\phi$ is a circle on the complex plane, where $f(m)$ determines the centre and $g$ – the radius. When $m \to \pm\infty$, $|f(m)| \to \pm\infty$. Thus the circle is far away from the origin and the winding number must be 0. Therefore the system is trivial when $m \to \pm\infty$. Thus the topological phases are between $m_{2i-1}$ and $m_{2i}$. Therefore, the number of topological phases are indeed equal to half of the number of $m_i$. Thus Eq. (8) is proved.

## TRANSITION POINTS IN AIII

To find transition points in class AIII the formalism of $q$ matrix should be somewhat modified. Note that this time the transitions may not occur at $\phi = 0$ or $\phi = \pi$ but at a generic $\phi$.

To solve for $\det Q = 0$, $f(m)$ and $g\exp(i\phi)$ should have the same phase, thus $\phi$ should be $\phi_m$ or $\phi_m + \pi$, where

$$\phi_m = \text{Arg}(f(m)) - \text{Arg}(g). \tag{10}$$

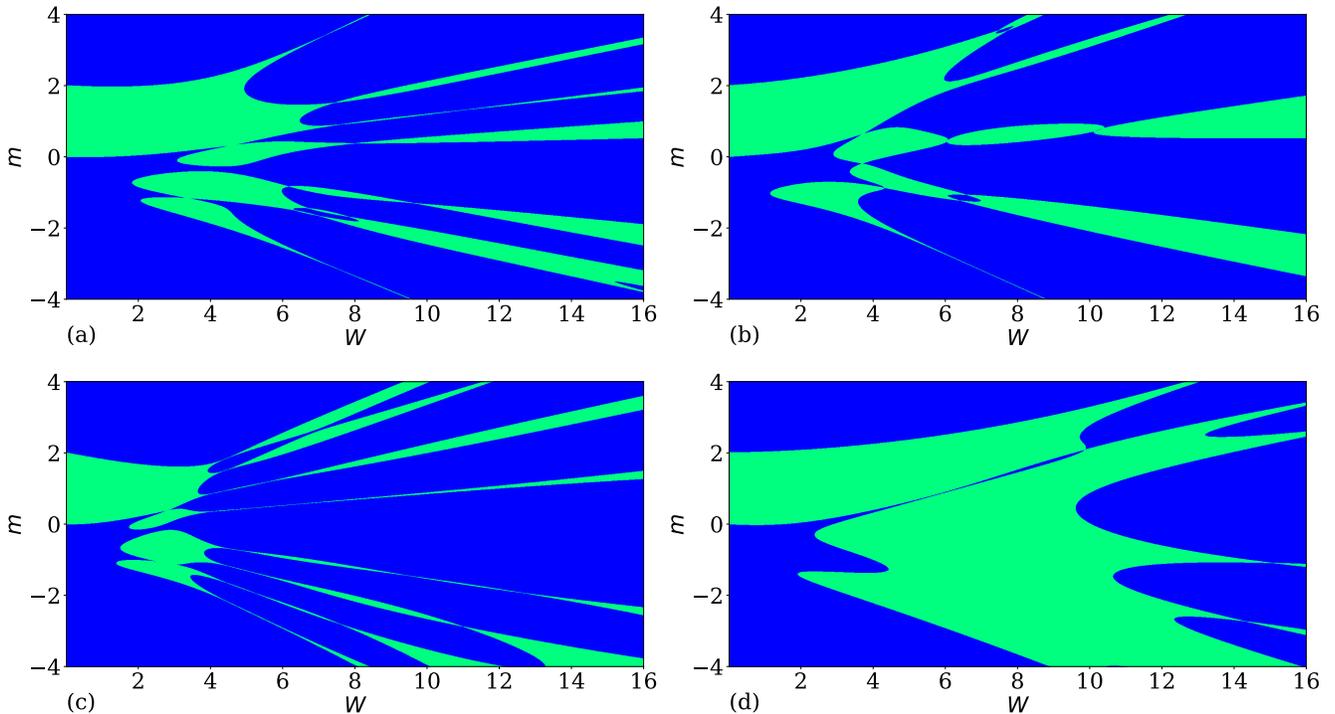

FIG. 1: Phase diagrams of model with longer ranges of hopping. The green region is $\nu = 1$, while the blue region is $\nu = 0$. The presence of topological islands illustrates its generality.

Thus,

$$f(m) - ge^{i\phi_m} = 0 \tag{11}$$
$$f(m) + ge^{i\phi_m} = 0 \tag{12}$$

Taking the complex conjugate of the second equation and multiplying the first equation,

$$f(m)f^*(m) - gg^* = 0 \tag{13}$$

A $2N \times 2N$ matrix $P$ needs to be constructed, where the diagonal items are $Ww_i$ and $W'w_i'^*$, the sub-diagonal and the upper right corner items are $1+W'w_i'$ and $1+W'w_i'^*$. In this way,

$$\det(P - (-m)I) = f(m)f^*(m) - gg^* = 0. \tag{14}$$

Thus all the transition points are real eigenvalues of $P$. This allows one to easily find all the transition points.

Note that when there are few transition points, one can directly solve zeros of $\log(|f(m)|/|g|)$ since at the transitions, $|f(m)| = |g|$.

## MODEL WITH LONGER RANGES OF HOPPING

Topological islands is a general feature in topological disordered systems. It is not limited to the specific model discussed in the main text. They are formed due to the properties of real zeros of the random polynomial $p(m,\phi) = \det Q(m,\phi)$, demonstrating its generality across various topological disordered systems. Topological islands exists as long as there are many real zeros.

To illustrate the generality of the topological islands, we consider here adding longer hopping terms

$$\sum_{j=1}^{N} t_j' c_{j,B}^\dagger c_{j+2,A} + h.c. \tag{15}$$

to the model discussed in the main text. Here

$$t_j' = 1 + W'w_j'', \tag{16}$$

where $W'$ is the hopping strength and $w_j''$ is random variables uniformly drawn from the box $[-1/2, 1/2]$.

Four typical phase diagrams are shown here, see Fig. 1, illustrating the generality of topological islands.


\fi

\end{document}